\documentclass[final,english]{elsarticle}
\usepackage[T1]{fontenc}
\usepackage[latin9]{inputenc}
\usepackage{varioref}
\usepackage{units}
\usepackage{amsmath}
\usepackage{amssymb}
\usepackage{graphicx}
\usepackage{esint}
\usepackage{caption}
\usepackage{subcaption}
\pdfpageheight\paperheight
\pdfpagewidth\paperwidth

\journal{Elsevier}
\usepackage{upgreek}
\usepackage[bookmarks,bookmarksopen,bookmarksdepth=6]{hyperref}
\usepackage{cases}
\usepackage{url}
\usepackage{lineno}
\usepackage{a4wide}
\usepackage{newtxtext,newtxmath,amsmath}
\makeatother

\usepackage{babel}
\begin{document}

\title{On the weighting field of irradiated silicon detectors}
%\title{}
\author[]{J.~Schwandt\corref{cor1}}
\author[]{R.~Klanner}

\cortext[cor1]{Corresponding author. Email address: Joern.Schwandt@desy.de.}

\address{ Institute for Experimental Physics, University of Hamburg,
 \\Luruper Chaussee 149, D\,22761, Hamburg, Germany.}

%\pdfbookmark[5]{Abstract}{abstract}}

\begin{abstract}
% \label{sect:Abstract}

 The understanding of the weighting field of irradiated silicon sensors is essential for calculating the response of silicon detectors in the radiation environment at accelerators like at the CERN LHC.
 Using 1-D calculations of non-irradiated pad sensors and 1-D TCAD (Technology Computer-Aided Design) simulations of pad sensors before and after irradiation, it is shown that the time-dependence of the weighting field is related to the resistivity of low field regions with ohmic behaviour in the sensor.
 A simple formula is derived, which relates the time constant of the time-dependent weighting field, $\tau $, with the resistivity and the extension of the low-field region for pad detectors.
 As the resistivity of irradiated silicon increases with fluence and finally reaches the intrinsic resistivity, $\tau $ becomes much larger than the charge-collection time and the weighting field becomes essentially independent of time.
 The TCAD simulations show that the transition from a time-dependent to a time-independent weighting field occurs at a neutron-equivalent fluence of $ \approx 5 \times 10^{12}$~cm$^{-2}$ for a $200~\upmu$m thick pad diode operated at 40~V and $- 20~^\circ$C.
 It is therefore concluded that the use of a time-independent weighting field calculated with the same method as for a fully-depleted non-irradiated sensor is also appropriate for the simulation of highly irradiated silicon sensors.
% For the analytic calculation of the non-irradiated pad sensor the close relation between the time dependence of the weighting field and the frequency dependence of the capacitance is shown.

\end{abstract}


\maketitle
% \tableofcontents
%--- \newpage
 \pagenumbering{arabic}

%\newpage

\section{Introduction}
 \label{sect:Introduction}
% Here the introduction
 The \emph{weighting field} describes the electromagnetic coupling of a charge  to an arrangement of conducting electrodes.
 It is used to calculate the signal currents induced in the readout electrodes by charges moving in a detector.
 The weighting field has first been introduced by Shockley and Ramo~\cite{Shockley:1938, Ramo:1939} to describe the signal generation in vacuum tubes.
 They also presented an elegant method of calculating weighting fields for different electrode arrangements.
 In Ref.~\cite{Cavalleri:1972} the method was extended to the presence of fixed space charges.
 In Ref.~\cite{Gatti:1982} the \emph{time-dependent weighting vector} has been introduced, to describe the situation when the electrodes are not kept at a fixed potential, but connected by linear impedance elements to ground or the power supply, and methods of its calculation were presented.
 The extension to non-linear media was presented in Ref.~\cite{Hamel:2008}, and to media with arbitrary conductivity and permittivity in Ref.~\cite{Riegler:2004}.
 A recent overview, which includes the calculation and discussion of the signal shape in partially depleted silicon pad detectors is given in Ref.~\cite{Riegler:2018}.
 It is found that a time-dependent weighting vector is required to describe this situation, and the time constant is given by $\tau = \rho \cdot \varepsilon _{Si}$ multiplied by a geometrical factor, which is bigger than one and is given by the depletion depth relative to the sensor thickness.
 The resistivity of the silicon is denoted $\rho $, and $\varepsilon _{Si} $ is the dielectric constant of silicon.
 A typical value of the resistivity of silicon before irradiation is $5~\mathrm{k}\Omega \cdot$cm, which  gives $\tau = 5$~ns times the geometric factor.

 This paper addresses the question how radiation damage by energetic particles influences the weighting field and thus the signal generation.
 The electric field in a highly irradiated silicon pad sensor exhibits a so-called \emph{double junction}: High-field regions at the two electrodes and a resistive low-field region in-between~\cite{Eremin:2002}.
 Thus the field distribution is not too different from the distribution in a partially depleted non-irradiated pad diode with implantations at both electrodes.
 For partially depleted non-irradiated pad diodes analytic results are available~\cite{Riegler:2018}.
 They show that a time-dependent weighting field is required to describe the signal shapes.
 Thus they can be considered a good test bed for understanding the detailed and complex TCAD (Technology Computer-Aided Design) simulations, and give some confidence that they also give precise results when applied to radiation-damaged sensors.
% As the resistivity of radiation damage is close to intrinsic, $\rho \approx 200~\mathrm{k}\Omega \cdot$cm at room temperature, the corresponding time constant $\tau = 200$~ns, and the need for a time-dependent weighting vector could be expected.

 In the following section, experimental data from capacitance-voltage (\emph{C--V}) measurements of pad diodes for different frequencies after irradiation to 1 MeV neutron equivalent fluences, $\Phi _{eq}$, by 24~GeV/c protons are shown.
 At low $\Phi _{eq}$-values a strong frequency dependence of the capacitance is observed, whereas at high $\Phi _{eq}$ it is independent of voltage and frequency and equal to the geometrical capacitance of the fully depleted pad sensor before irradiation.
 A similar behaviour is obtained from a model calculation of a partially-depleted non irradiated pad diode.
 This suggests that the frequency dependence of the capacitance of irradiated sensors is influenced by the presence of a resistive low-field region.
% is not directly related to the detailed properties of the radiation-induced states in the silicon band gap.
 However, the frequency at which the capacitance changes for non-irradiated sensors is beyond the 2~MHz maximum frequency of most capacitance bridges, and has not been observed experimentally.
 %Within this model the frequency dependence is the result of the finite resistivity of the non-depleted region.

 Next, the time-dependent weighting field from the TCAD simulation is compared to the results of the analytical calculation for the non-irradiated diode.
 Apart from some expected minor differences, the agreement is good.
 In addition, the analytical model is used to derive a relation between the time constant of the time dependence of the weighting field and the depletion width and  resistivity of the non-depleted region.

 Finally, the \emph{Hamburg Penta-Trap Model}, HPTM~\cite{Schwandt:2018} is introduced, which is used  for the TCAD simulations of the electric field and the weighting field of pad diodes irradiated to different fluences.
 The fluence dependence of the electric field and the weighting field is presented, and used to extract the resistivity of the low-field region as a function of the fluence with the help of the formula derived from the analytic calculation of the non-irradiated pad diode.
 The results confirm that the time dependence of the weighting field is mainly determined by the resistivity and extension of the low field region of the irradiated sensor.

 \section{\emph{C--V--f} results: Data and simulations}
  \label{sect:CVf}

 Before discussing in detail the weighting field of non-irradiated and irradiated sensors, capacitance-voltage (\emph{C--V}) results are presented for a pad sensor of an area $ A = 0.5 \times 0.5$~cm$^2$ with the parameters given in Table~\ref{tab:Diode}.
 In the analysis the parallel capacitance, $C_p$, is used.
 From the admittance, $Y$, calculated for an electrical model, $C_p$ is obtained using $Y = 1/R_p + i~ \omega ~C_p$, with the parallel resistance $R_p$ and the angular frequency $\omega = 2 \pi f$.
 For the measurements the capacitance meter applies the voltage $V + V_{AC} \cdot \sin(\omega ~t)$ to the device under test, measures the amplitude of the current and the phase difference between current and voltage, and calculates the corresponding values of $1/R_p (V, f) $ and $C_p (V, f)$.

 \begin{table} [!ht]
  \centering
   \begin{tabular}{c|c|c|c|c|c}
    $d$ & $N_p$ & $N_{n^+}$ & $d_{n^+}$ & $N_{p^+}$ &  $d_{p^+}$ \\
    $[\upmu $m] & [$10^{12}$~cm$^{-3}$] & [$10^{19}$~cm$^{-3}$] & $[\upmu $m] & [$10^{19}$~cm$^{-3}$] & $[\upmu $m] \\     \hline
    200 & 3.85 & 1.2 & 2.4 & 1.0 & 2.4 \\
%    \hline
   \end{tabular}
    \caption{Pad-diode parameters used for the TCAD simulations:
    $d$ is the mechanical Si-thickness of the pad diode,
    $N_p$ the boron density of the silicon bulk,
    $d_{n^+}$ and $N_{n^+}$ the depth and dopant density of the phosphorus implant, and
    $d_{p^+}$ and $N_{p^+}$ the depth and dopant density of the backside boron implant.
    \label{tab:Diode}}
  \end{table}

 Before irradiation, the measurements agree with the expectations: Below the full depletion voltage, $V_{fd} \approx 120$~V,  $1/C_p^2$ depends linearly on voltage, with the parallel capacitance $C_p = (\varepsilon _{Si} \cdot A )/w(V)$ and the depletion depth $w(V) = \sqrt{ 2 \varepsilon _{Si} \cdot (V + V_{bi}) / (q_0 \cdot N_p)}$.
 Above $V_{fd}$, $C_p \approx (\varepsilon _{Si} \cdot A )/d$.
 For the frequency range of the measurements, $f = 100~\mathrm{Hz} - 2$~MHz, the results do not depend on $f$.
 The dielectric constant of silicon is denoted $\varepsilon _{Si}$, the elementary charge $q_0$, and the built-in voltage $V_{bi} \approx 0.8$~V.

\begin{figure}[!ht]
   \centering
    \includegraphics[width=0.5\textwidth]{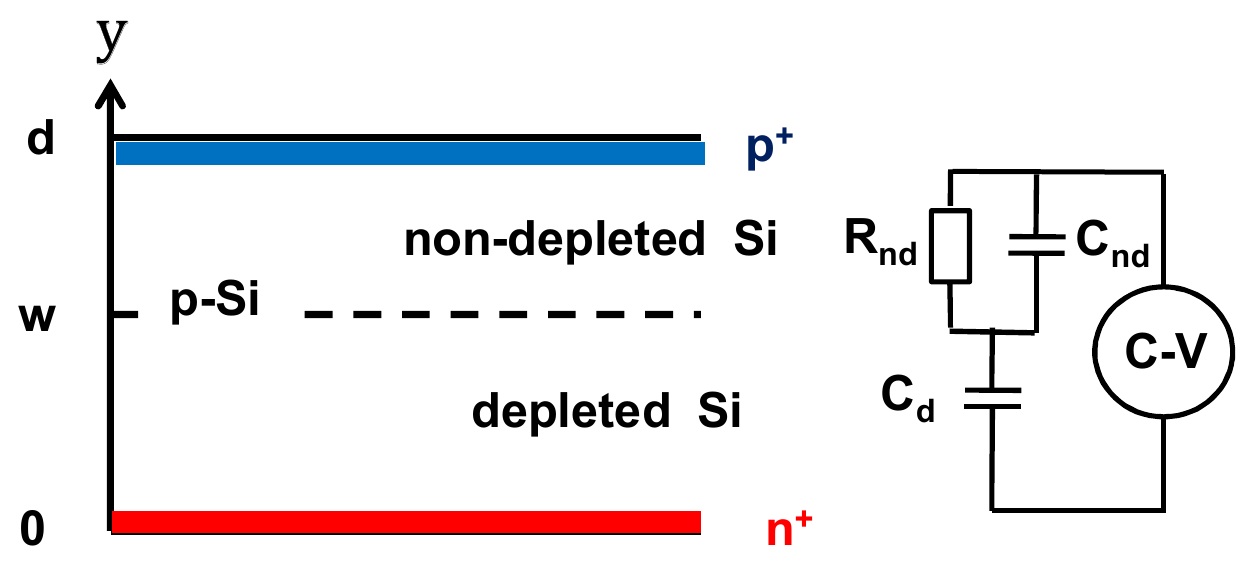}
   \caption{Left: Schematic cross-section and symbols used for the simulation of the pad diode.
   Right: Electrical model for the simulation of the partially depleted diode. $C_d$ refers to the capacitance of the depleted region, and $C_{nd}$ and $R_{nd}$ to the capacitance and the resistance of the non-depleted region.}
  \label{fig:Sensor}
 \end{figure}

 Next, the $C_p (V, f)$-dependence is calculated using a simple model.
 The left side of Fig.~\ref{fig:Sensor} shows a schematic cross-section of the sensor.
 The $n^+p$ diode is at $y = 0$, and the backside $p^+$ implant at $y = d$.
 The dashed line at $ y = w$ indicates the boundary between the depleted and the non-depleted regions of a non-irradiated pad diode.
 The right side of the figure shows the electrical model used to simulate the expected \emph {C--V--f} dependence.
 The capacitance of the depletion region $C_d$, and the capacitance $C_{nd}$  and the resistance $R_{nd}$  of the non-depleted region are:
 \begin{equation}\label{equ:Elpar}
   C_d = (\varepsilon _{Si} \cdot A )/w(V), \hspace{5mm}
   C_{nd} = (\varepsilon _{Si} \cdot A )/(d - w(V)), \hspace{5mm}
   R_{nd} = \rho \cdot (d - w(V))/ A ,
 \end{equation}
 with the resistivity of the non-depleted silicon, $\rho = 1/(q_0 \cdot \mu _h \cdot N_p )$.

\begin{figure}[!ht]
   \centering
   \begin{subfigure}[a]{0.5\textwidth}
    \includegraphics[width=\textwidth]{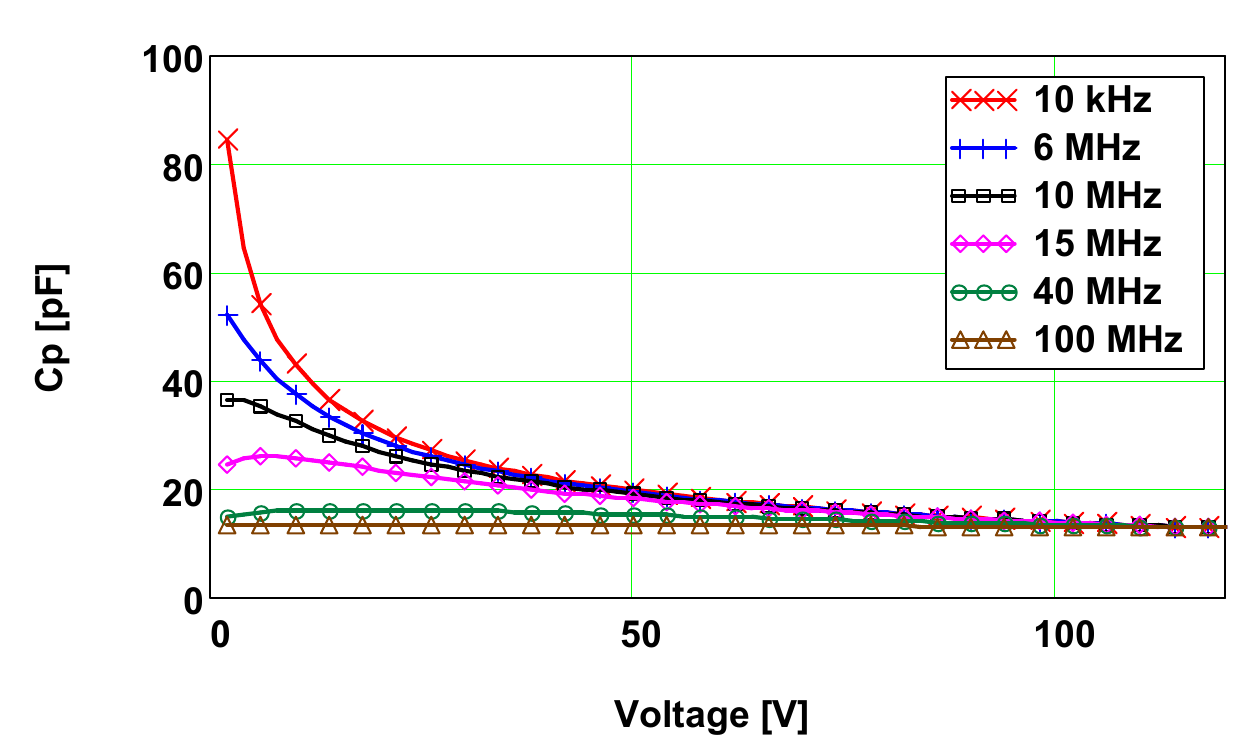}
    \caption{ }
    \label{fig:CVfnon}
   \end{subfigure}%
    ~
   \begin{subfigure}[a]{0.5\textwidth}
    \includegraphics[width=\textwidth]{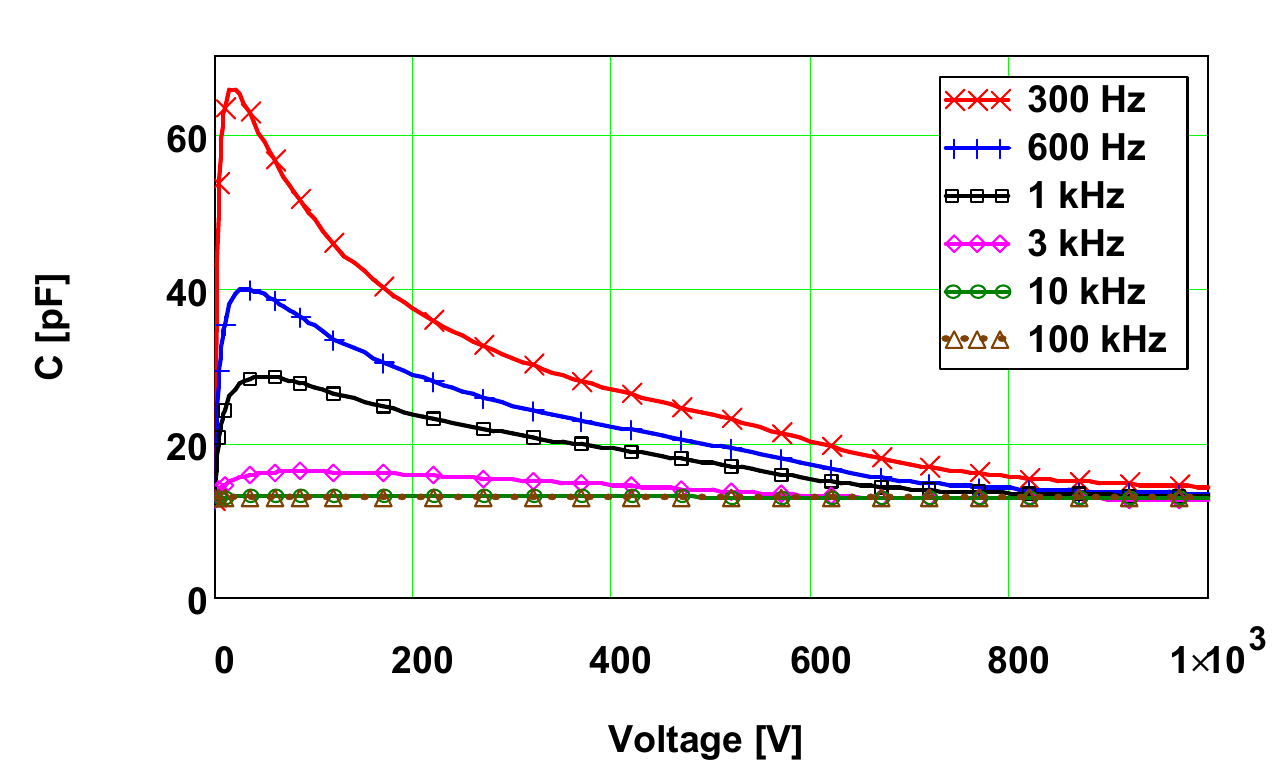}
    \caption{ }
    \label{fig:CVfirr}
   \end{subfigure}%
   \caption{ (a) Simulated parallel capacitance-voltage dependence at different frequencies for the non-irradiated $n^+p$~pad diode of area $0.5 \times 0.5$~cm$^2$ and thickness $200~\upmu$m.
   (b) Parallel capacitance-voltage dependence measured at different frequencies for the $n^+p$~pad diode irradiated by 24 GeV/c protons to $\Phi _{eq} = 6 \times 10^{15}$~cm$^{-2}$. The measurements were performed at $- 30~^\circ$C.
    }
  \label{fig:CVf}
 \end{figure}

 The electric model gives for the parallel capacitance:
% From the admittance $ Y = 1/R_p + i~\omega~C_p$ of the electrical model the parallel capacitance is obtained
 \begin{equation}\label{equ:Cp}
   C_p(V,\omega) = C_d \cdot \frac{1+\omega ^2 \cdot \tau _{nd} \cdot (\tau _{nd} + \tau _{d}) } {1 + \omega ^2 \cdot (\tau _{nd} + \tau _{d}) ^2 },
 \end{equation}
 with $\tau _{nd}(V) = C_{nd}(V) \cdot R_{nd}(V)$ and $\tau _{d}(V) = C_{d}(V) \cdot R_{nd}(V)$.
 The calculated voltage dependence of $C_p$ for different frequencies $f$ is shown in Fig.~\ref{fig:CVfnon}.
 Up to $f \approx 3$~MHz, the voltage dependence is: $C_p(V) = (\varepsilon _{Si} \cdot A )/w(V)$.
 For $f = 3 - 50$~MHz the value of $C_p$ decreases at low voltages, and finally reaches the constant value $C_p =  (\varepsilon _{Si} \cdot A )/d$ for $f \gtrsim 50$~MHz.
 At these high frequencies the entire AC-current of the capacitance meter flows through $C_{nd}$  and $C_{d}$, and the capacitance is $C_p = (1/C_{nd} +1/C_{d})^{-1} = (\varepsilon _{Si} \cdot A )/d$, the geometrical capacitance of the pad diode.

 Fig.~\ref{fig:CVfirr} shows the measured voltage dependence $C_p (V)$ at different frequencies for the pad sensor irradiated by 24~GeV/c protons to a neutron-equivalent fluence $\Phi _{eq} = 6 \times 10^{15}$~cm$^{-2}$.
% As shown in Fig.~\ref{fig:CVf}, the situation after irradiation is completely different.
% The figure shows the voltage dependence of the parallel capacitance, $C_p (V)$, at different frequencies for the pad sensor irradiated by 24~GeV/c protons to $\Phi _{eq} = 6 \times 10^{15}$~cm$^{-2}$ .
 For the conversion to $\Phi _{eq}$ a hardness factor of 0.62 is used.
 The estimated uncertainty of the fluence is $\pm 10~$\%.
 At low frequencies, a strong dependence of $C_p$ on $f$ is observed.
 However, for $f \gtrsim 10$~kHz, $C_p$ is constant and equal to the geometrical capacitance of the pad diode.
 The corresponding frequency for the partially depleted pad diode before irradiation is 100~MHz.

 Frequently the observation of a frequency-dependent $C_p$ is assumed to be caused by the emission-capture times of the radiation-induced states in the silicon band gap, which can change their charge states only up to a certain frequency.
 In Refs.~\cite{Oldham:1972, Beguwala:1974, Dabrowski:1989, Li:1991, Borchi:1998} $C_p(V,f)$ is  used to extract information about the properties of  radiation induced states.
 However, as demonstrated by Fig.~\ref{fig:CVfnon}, also non-irradiated sensors, operated below the full depletion voltage, exhibit a frequency-dependent $C_p$, which is caused by the finite resistivity of the non-depleted silicon.
 As discussed in Sect.~\ref{sect:Ewirr}, the resistivity of the non-depleted silicon increases by several orders of magnitude with irradiation.
 As a result the frequency dependence of $C_p$ calculated with Eq.~\ref{equ:Cp}, is strongly influenced by the change of the resistivity due to radiation damage.
 In Refs.~\cite{Campbell:2001, Campbell:2002} the dependence of the resistivity on fluence and temperature is used to derive a frequency-temperature scaling law of the $C-V$~characteristics for irradiated silicon detectors.
 This scaling law is frequently used to estimate the depth of the depleted region in irradiated silicon detectors.

 As the frequencies relevant for the charge collection are $\gtrsim 10$~MHz (approximately the inverse of the charge collection time of $\approx 10$~ns), the discussion above suggests that before irradiation a time-dependent weighting field is required to describe the pulse shapes from partially depleted sensors.
 However, after irradiation the weighting field is expected to be time-independent and equal to the weighting field of the fully depleted sensor before irradiation.
 Addressing this question by detailed simulations is the main aim of this paper.

% \begin{figure}[!ht]
%   \centering
%    \includegraphics[width=0.6\textwidth]{FigCVf.pdf}
%   \caption{Parallel capacitance-voltage measurement results at different frequencies for the $n^+p$~pad diode of area $0.5 \times 0.5$~cm$^2$ irradiated by 24 GeV/c protons to $\Phi _{eq} = 6 \times 10^{15}$~cm$^{-2}$. The measurements were performed at $- 30~^\circ$C.}
%  \label{fig:CVf}
% \end{figure}

 \section{Time-dependent weighting vector and weighting field}
  \label{sect:Ew}

 The time-dependent weighting vector ${\overrightarrow{W}}(\vec{r},t)$  has been introduced in Ref.~\cite{Gatti:1982} to describe the signal in detectors when the electrodes are not grounded.
 In Ref.~\cite{Riegler:2004} it is shown that the same formalism can be used if the material between the electrodes is conducting.
% For ${\overrightarrow{W}}(\vec{r},t)$,
% a charge $Q_0$ created at $t=0$ moving on the trajectory $\vec{r}(t)$ with the velocity $\vec{v}(t)$ induces on the grounded electrode at the time $t$ the current
 The current induced in an electrode at constant potential by a charge $Q_0$ created at $t=0$ moving on the trajectory $\vec{r}(t)$ with the velocity $\vec{v}(t)$  is
 \begin{equation}\label{equ:It}
%  I(t) = Q_0 \cdot  \int_0 ^t {\vec{v}}(\vec{r}(t'),t - t') \cdot \overrightarrow{W}(\vec{r}(t'))~\mathrm{d}t'.
    I(t) = Q_0 \cdot  \int_0 ^t {\overrightarrow{W}}\big(\vec{r}(t'),t - t'\big) \cdot \vec{v}(t')~\mathrm{d}t'.
 \end{equation}
% The time-dependent weighting field is defined by the relation
%\begin{equation}\label{DefEw}
%  \overrightarrow{E_w}(\vec{r},t) = \int_0 ^t {\overrightarrow{W}}(\vec{r},t')~\mathrm{d}t',
%\end{equation}
% and the time-dependent weighting vector is obtained from $\overrightarrow{E_w}(\vec{r},t)$ by
  The time-dependent weighting vector is related to the time-dependent weighting field  $\overrightarrow{E_w}(\vec{r},t)$ by
\begin{equation}\label{equ:Wyt}
 {\overrightarrow{W}}(\vec{r},t) =  \frac{\partial} {\partial t} \overrightarrow{E_w}(\vec{r},t).
\end{equation}
 The weighting field can be calculated in the following way:
 First the electrical field $\overrightarrow{E}(\vec{r})$ for a sensor biased with the voltage $V$ is calculated.
 Then, a voltage step $V_0 \cdot \Theta(t)$ with a small $V_0$~value is applied to the readout electrode, and the time-dependent field $\overrightarrow{E_t}(\vec{r},t)$ is calculated, to obtain
 \begin{equation}\label{equ:Ew}
   {\overrightarrow{E_w}}(\vec{r},t) = \frac {1} {V_0} \cdot \Big(\overrightarrow{E_t}(\vec{r},t) - \overrightarrow{E}(\vec{r}) \Big).
 \end{equation}
 $\Theta(t)$ is the Heaviside step function.
 For numerical calculations $V_0 \cdot \Theta (t)$ has to be replaced by a fast voltage ramp from 0 to $V_0$.
 Note that the electric field has units voltage/length, the weighting field 1/length, and the weighting vector 1/(length$\cdot$time).

 ${\overrightarrow{E_w}} (\vec{r},t)$ has two terms:
 A quasi-instantaneous step,  and a time-dependent term, which accounts for the flow of charges in the conductive material of the sensor and/or the charge flow through the connections of the electrodes to the ground and the power supply.
 ${\overrightarrow{E_w}} (\vec{r},0^+)$\footnote{$~0^-$ denotes the time just before $t = 0$, and $0^+$ just after $t = 0$.}
 is equal to the \emph{geometric weighting field}, $\overrightarrow{E}_{geom}(\vec{r})$, which is obtained by just considering the electrodes and ignoring the effects of the conductive material in the sensor.
 The time constant of the quasi-instantaneous term is short, but finite, as electric fields propagate with the speed of light.
 For a sensor of a thickness of $200~\upmu$m it is of the order of 10~fs, can be considered instantaneous and thus ignored.
 When calculating $\overrightarrow{W}(\vec{r},t)$ using Eq.~\ref{equ:Wyt}, the instantaneous term $\overrightarrow{E}_{geom}(\vec{r}) \cdot \Theta (t)$ gives the instantaneous contribution $\overrightarrow{E}_{geom}(\vec{r}) \cdot \delta (t)$.
 From Eq.~\ref{equ:It} it follows, that, in the absence of an additional time-dependent term, the induced current in the readout electrode from a charge $Q_0$ moving with the velocity $\vec{v}(\vec{r}(t))$ at time $t$ is
 \begin{equation}\label{equ:Igeom}
  I(t) = Q_0 \cdot \vec{v}(\vec{r},t) \cdot \overrightarrow{E}_{geom}(\vec{r}) \cdot \Theta(t),
 \end{equation}
 which is the formula usually used to calculate the signals in detectors.

 From Eq.~\ref{equ:It} it also follows that $\overrightarrow{E_{w}}(\vec{r},t)\cdot\mathrm{d}\vec{r}$ describes the time dependence of the charge induced in the readout electrode by a unit charge which has moved from $\vec{r}$ to $\vec{r} + \mathrm{d}\vec{r}$ at $t=0$.
 The corresponding current is $\overrightarrow{W}(\vec{r},t)$.

 \section{Weighting field of the non-irradiated pad diode}
  \label{sect:Ewnonirr}

 In this section 1-D TCAD simulations are compared to the analytical calculation of an $n^+p$ pad diode.
 The schematic layout of the diode and the coordinate system used are shown in Fig.~\ref{fig:Sensor}, and the parameters are given in Table~\ref{tab:Diode}.
 The silicon bulk is doped with boron and does not contain additional levels in the silicon band gap due to irradiation or impurities.
 The electrode at $y = d$ is connected to ground and the electrode at $y=0$ is biased to the voltage $V$.
 Given that a 1-D problem is analysed, only the $y$-components of the vectors introduced in Sect.~\ref{sect:Ew} are relevant, and vector signs are omitted in the following.

 \subsection{Analytical calculation}
  \label{subsect:analyical}

 In Ref.~\cite{Riegler:2018} the Laplace-transform technique is used to calculate the weighting vector, $ W(V,y,t)$, and the signal current $I(t)$ for holes and electrons produced at $t = 0$ with different $y$-distributions.
 In the following calculation the electrical model shown in Fig.~\ref{fig:Sensor} is used to derive an analytical expression for the time-dependent weighting field $E_w(V,y,t)$ and the weighting vector $W(V,y,t)$.
 For the calculation, at $t = 0$ a voltage step of height $V_0$ is added to the bias voltage $V$ at $y = 0$.

 For voltages above the full-depletion voltage, $V \geq V_{fd}$,  the \emph{geometric weighting field}, $E_{geom} = 1/d$.
 As there are no free charges in the depleted sensor, the resistivity is infinite, and $E_w(V,y,t) = 1/d $ for $ t > 0$ independent of $V,~y$ and $t$.
 To simplify the formulae, $d$ is used for the sensor thickness, instead of $d' \approx d - (d_{n^+} + d_{p^+})$, which takes into account the field-free regions of the highly doped regions.
 For the figures, in which the analytical results are compared to the TCAD simulations, this small difference is taken into account.

 For $V < V_{fd}$ the weighting field immediately after the voltage step, $E_w(V,y,0^+) = E_{geom} = 1/d$, and the voltage at the depletion depth $w(V)$ is $V_w(t=0^+) = V_0 \cdot C_{nd} / ( C_{nd} +C_{d} )$.
 The differential equation for $V_w(t \geq 0^+)$ is obtained from current conservation:
 The current in the depleted region (right-hand side of the equation) is equal to the current in the non-depleted region (left-hand side)
 \begin{equation}\label{equ:Icons}
   \frac{V_w(t)} {R_{nd}} + C_{nd} \frac{\mathrm{d}V_w(t)} {\mathrm{d}t} = C_{d} \frac{\mathrm{d}\big(V_0 - V_w(t)\big)} {\mathrm{d}t},
 \end{equation}
 and the differential equation is
 \begin{equation}\label{equ:dVwdt}
   \frac{\mathrm{d}V_w(t)} {\mathrm{d}t} + \frac{V_w(t)} {R_{nd} \cdot (C_{nd} + C_d)} = 0.
 \end{equation}
 For the given initial conditions the solution is:
 \begin{equation}\label{equ:Vwt}
   \frac{V_w(t)} {V_0} = \frac{C_{d}} {C_{nd} + C_d}~ e^{-t /\big(R_{nd} \cdot (C_{nd} + C_d)\big)} =
   \frac{d-w} {d}~ e^{- t / \tau}.
 \end{equation}
 With the relations given in Eq.~\ref{equ:Elpar}, one finds
  \begin{equation}\label{equ:tau}
   \tau(V) = \varepsilon_{Si} \cdot \rho \cdot d/w(V) = \tau _r \cdot d/w(V),
  \end{equation}
 with the dielectric relaxation time $\tau _r = \varepsilon_{Si} \cdot \rho$.
 For the parameters given in Table~\ref{tab:Diode}  $\tau_r = 3.5$~ns at 20$~^\circ $C.
 For $V < V_{fd}$ and $t > 0$ the time-dependent weighting field is obtained by dividing $(V_0-V_w)/V_0$ by $w$ for $y < w$, and $V_w/V_0$ by $d-w$ for $y \geq w$:

   \begin{equation}\label{equ:Ewmodel}
    E_{w}(V,y,t) = \left\{
           \begin{array}{cc}
    \vspace{2mm}
             \frac{1} {w(V)} \big(1 - \frac{d - w(V)} {d}~ e^{-t/\tau(V)} \big) \cdot \Theta(t)
             & \hbox{\rm{for}\,\, $0 < y < w(V)$} \\
    \vspace{1mm}
             \frac {1} {d} ~ e^{-t/\tau(V)} \cdot \Theta(t)  & \hbox{\rm{for}\,\, $ w(V) \leq y < d$}. \\
           \end{array}
         \right.
  \end{equation}

 Using Eq.~\ref{equ:Wyt} gives

    \begin{equation}\label{equ:Wmodel}
    W(V,y,t) = \left\{
           \begin{array}{cc}
    \vspace{2mm}
             \frac{1} {d} \Big( \delta (t) + \frac{d - w(V)} {w(V) \cdot \tau(V)}~ e^{-t/\tau(V)} \Big)
             & \hbox{\rm{for}\,\, $0 < y < w(V)$} \\
    \vspace{1mm}
             \frac {1} {d} \Big( \delta (t) - \frac {1} {\tau(V)} ~ e^{-t/\tau(V)} \Big)  & \hbox{\rm{for}\,\, $ w(V) \leq y < d$}. \\
           \end{array}
         \right.
  \end{equation}
 $W$ is zero for $t < 0$.
 The result agrees with the result of Ref.~\cite{Riegler:2018}, which uses the Laplace transform.

\begin{figure}[!ht]
   \centering
   \begin{subfigure}[a]{0.5\textwidth}
   \includegraphics[width=\textwidth]{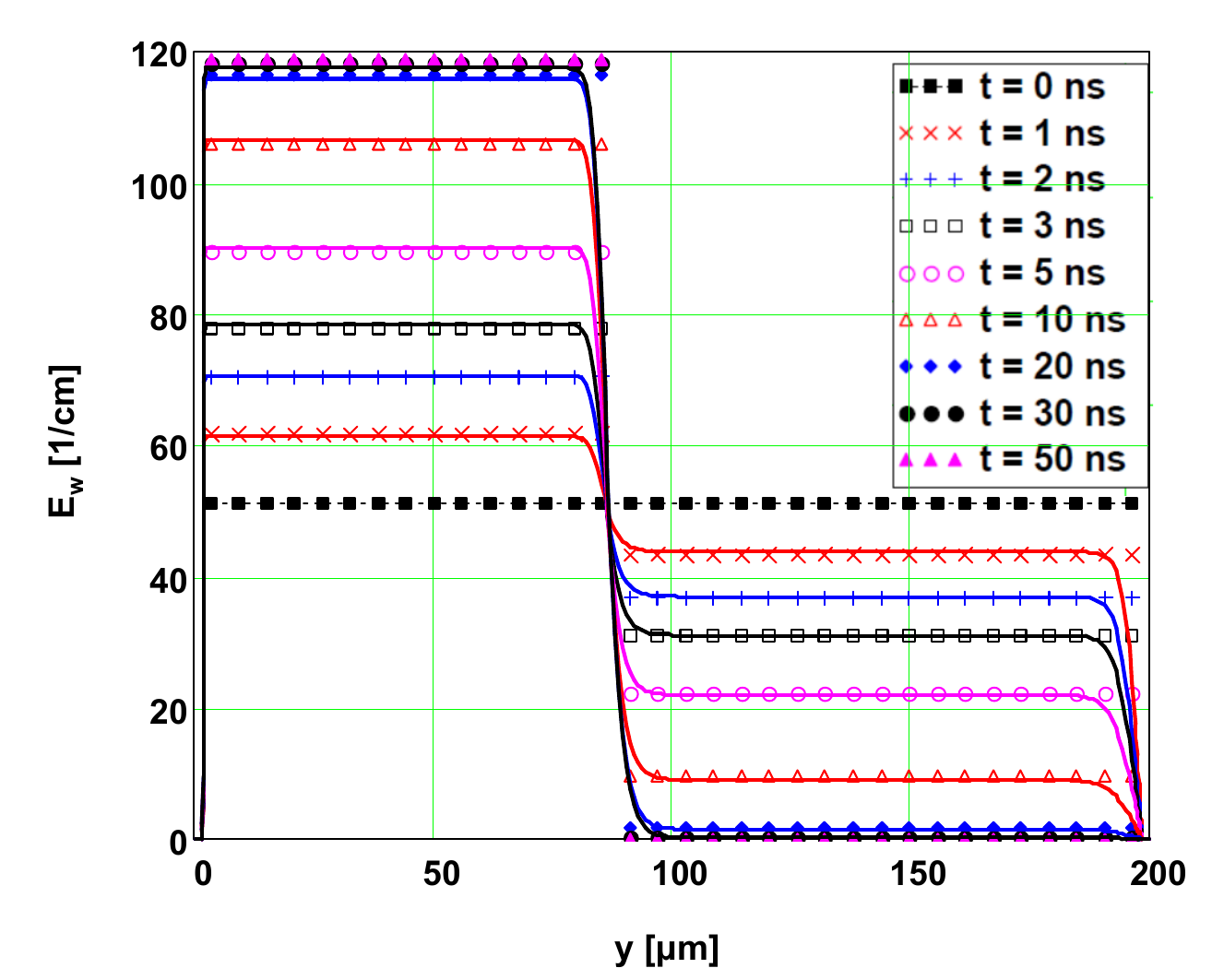}
     \caption{ }
    \label{fig:Etw-y-20V}
   \end{subfigure}%
    ~
   \begin{subfigure}[a]{0.5\textwidth}
    \includegraphics[width=\textwidth]{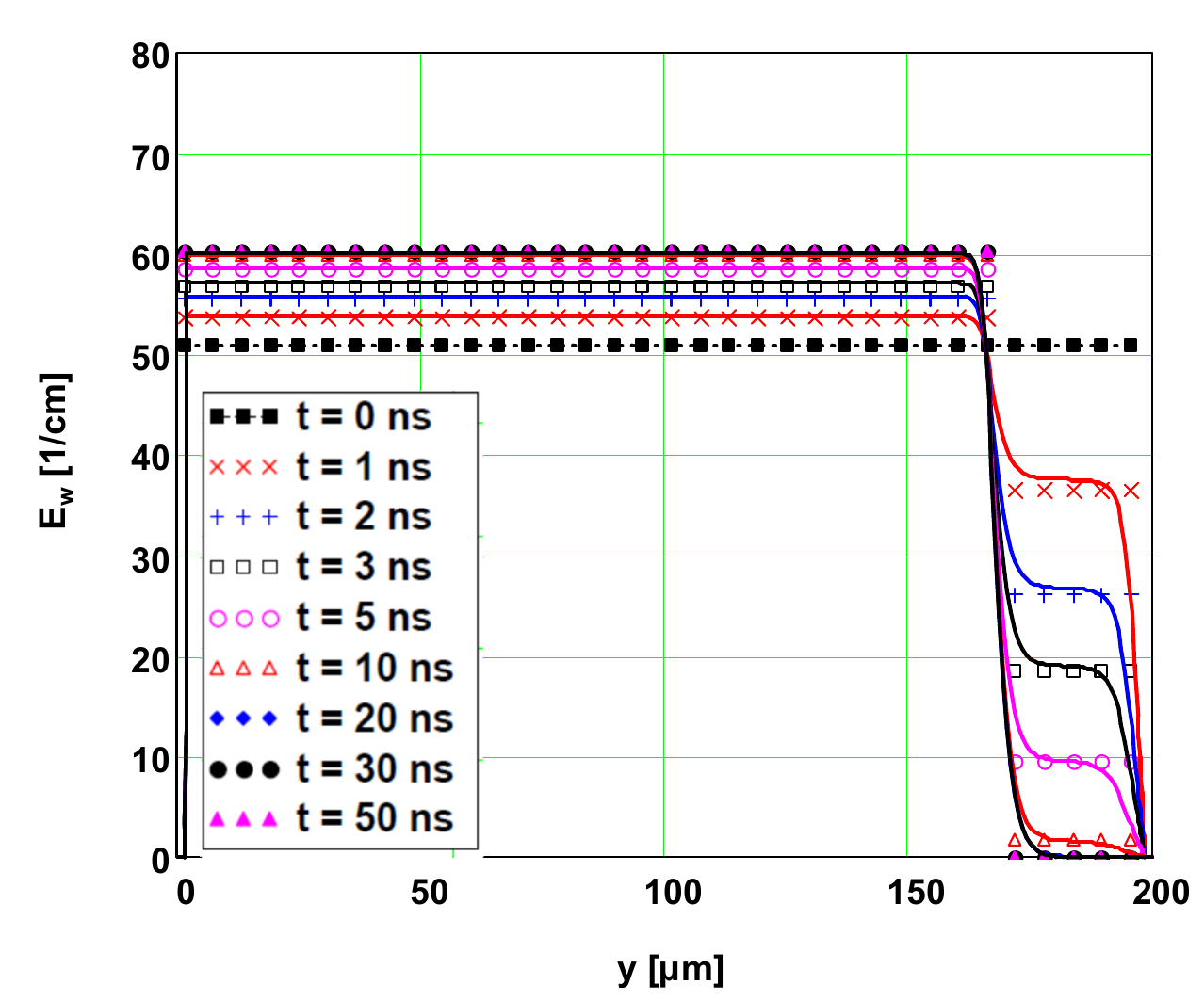}
    \caption{ }
    \label{fig:FigEtw-y-80V}
   \end{subfigure}%
   \caption{Comparison of the analytical calculation (symbols) and the TCAD simulation (lines) of the time-dependent weighting field, $E_{w}(V,y,t)$,  for the non-irradiated pad diode.
%   The crosses for $E_{w} = 50 $~cm$^{-1}$ correspond to $ t = 0$, the following crosses and lines correspond to $E_{w}(V,y,t)$ for 1~ns increments in $t$ up to $t = 50$~ns.
     (a) Results for $V = 20$~V, and
     (b) results for $V = 80$~V.
    }
  \label{fig:Etw-y}
 \end{figure}

\begin{figure}[!ht]
   \centering
   \begin{subfigure}[a]{0.5\textwidth}
    \includegraphics[width=\textwidth]{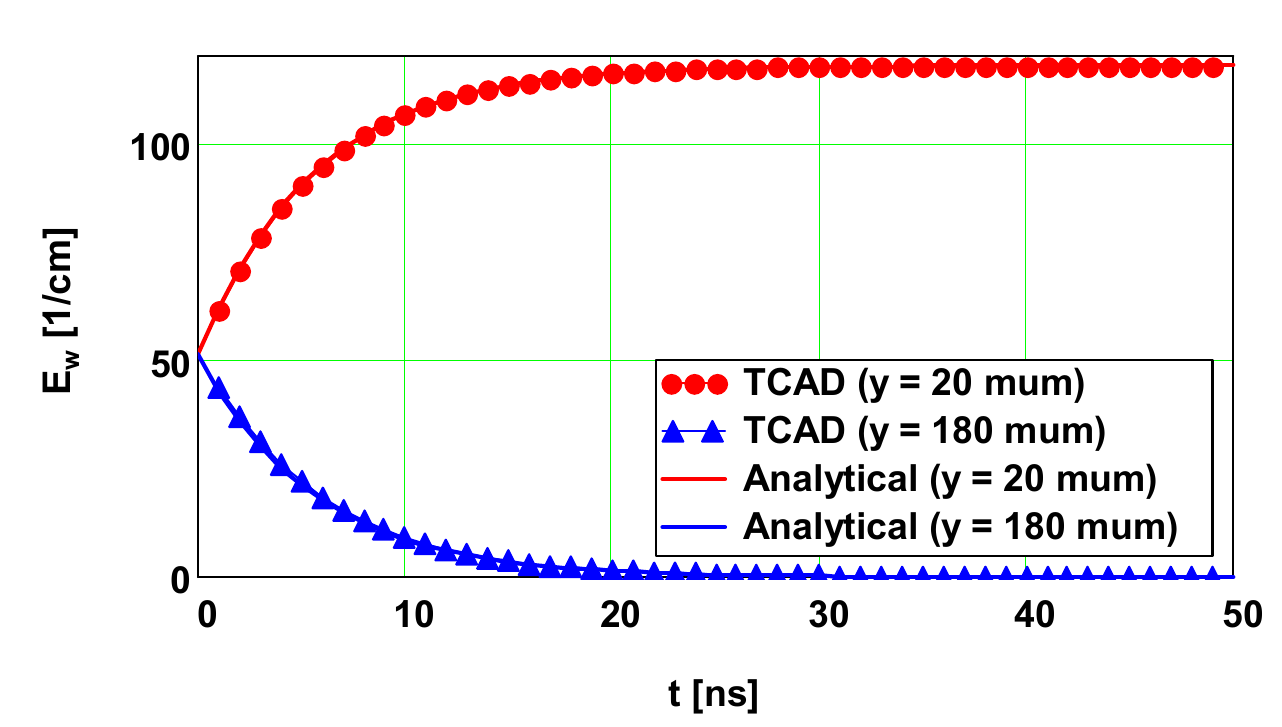}
    \caption{ }
    \label{fig:Etw-t-20V}
   \end{subfigure}%
    ~
   \begin{subfigure}[a]{0.5\textwidth}
    \includegraphics[width=\textwidth]{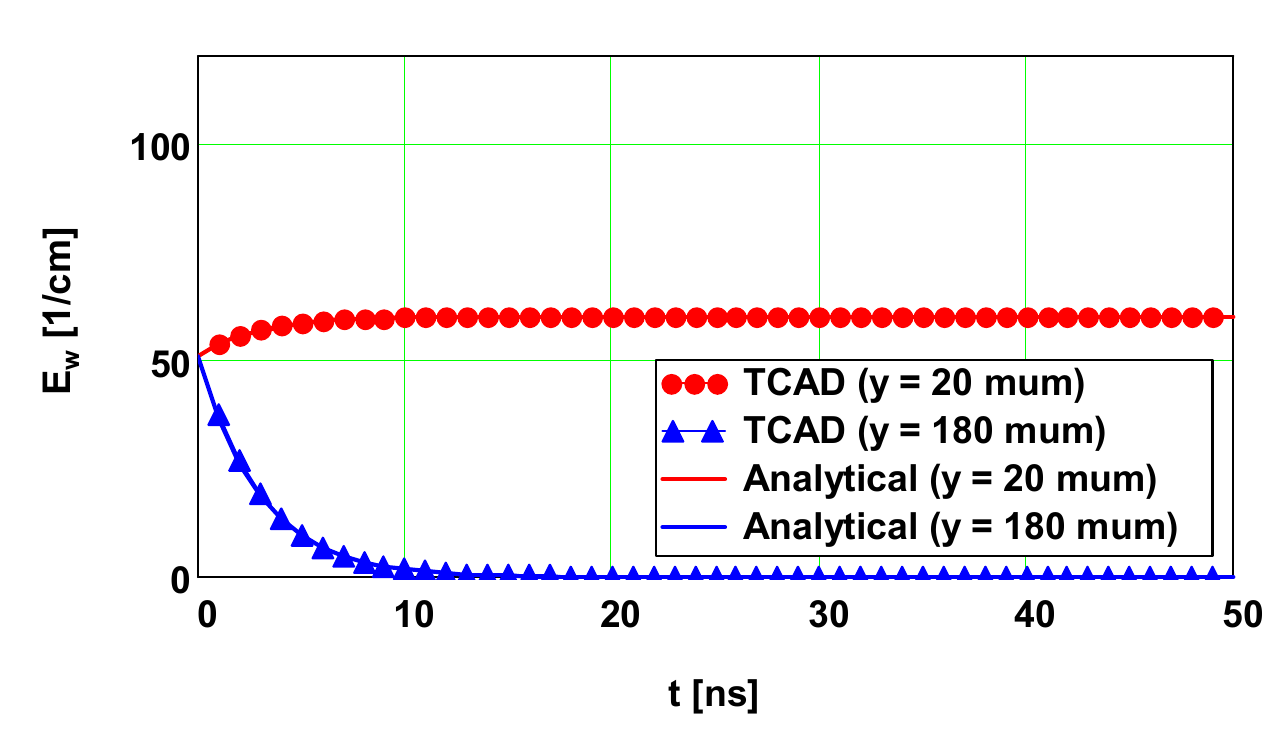}
    \caption{ }
    \label{fig:FigEtw-t-80V}
   \end{subfigure}%
   \caption{Comparison of the analytical calculation (lines) and the TCAD simulation (symbols) of the time-dependent weighting field for the non-irradiated pad diode at $y = 20~\upmu$m (depleted region) and at $y = 180~\upmu$m (non-depleted region).
     (a) Results for $V = 20$~V, and
     (b) results for $V = 80$~V.
    }
  \label{fig:Etw-t}
 \end{figure}

 Figs.~\ref{fig:Etw-y} and \ref{fig:Etw-t} show, in addition to the results of the TCAD simulation discussed ins Sect.~\ref{subsect:TCAD-nonirr}, the calculated $E_w(V,y,t)$ for $V = 20$~V and $V = 80$~V.
 Both voltages are below the full-depletion voltage $V_{fd} \approx 120$~V.
 For $t = 0^+$, $E_{w} \approx 1/d = 50$~cm$^{-1}$, independent of $y$.
 The figures show that $E_{w}$ increases as a function of $t$ to $1/w$ in the depletion region $0 < y < w$, and decreases to zero in the non-depleted region $w < y < d$.

 \subsection{TCAD simulation}
  \label{subsect:TCAD-nonirr}

\begin{figure}[!ht]
   \centering
    \includegraphics[width=0.6\textwidth]{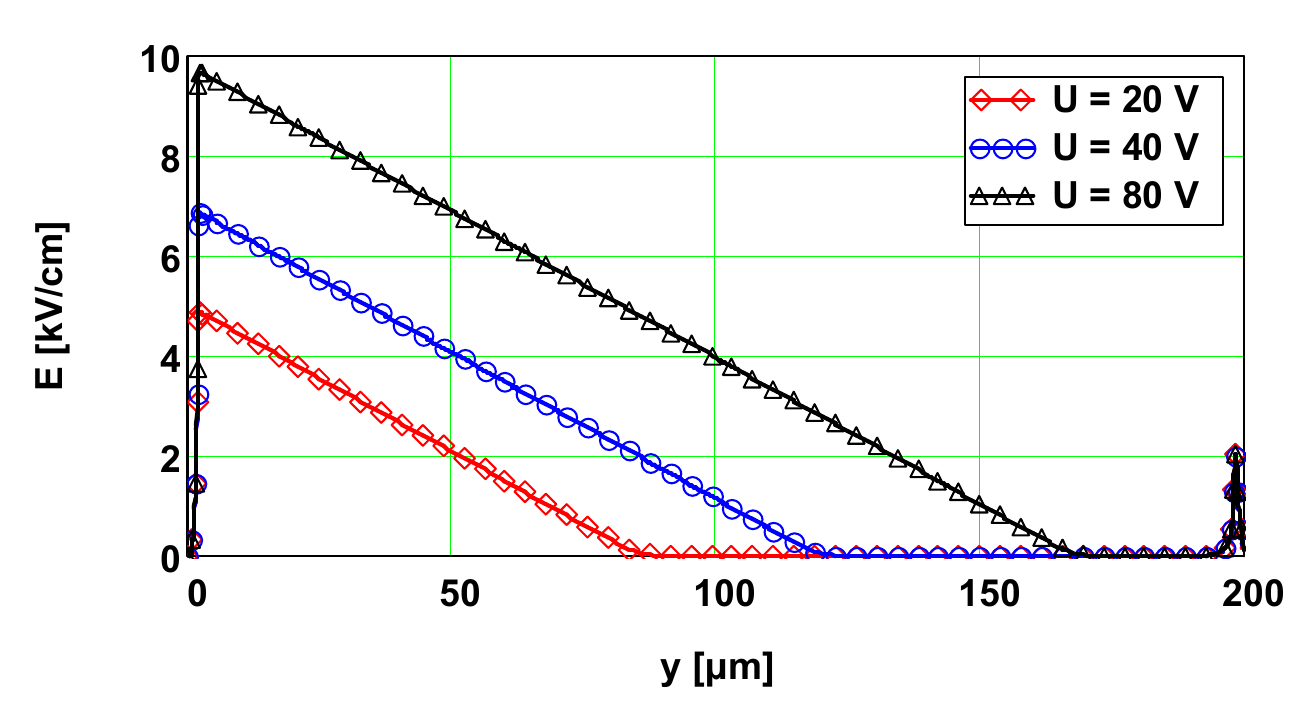}
   \caption{Electric field of the non-irradiated pad diode for $V = 20$, 40 and 80~V simulated using SYNOPSYS TCAD at $- 20~^\circ$C.
    Note the effects of the $n^+$-implant at $ y = 0$, and of the $p^+$-implant at $ y = 200~\upmu$m.}
  \label{fig:Efield-nonirr}
 \end{figure}

 With the pad-diode parameters of Table~\ref{tab:Diode} SYNOPSYS TCAD~\cite{TCAD} is  used to simulate the electric field, $E(V, y)$, and the weighting field $E_{w}(V,y,t)$.
 Fig.~\ref{fig:Efield-nonirr} shows $E(y)$ for $V = 20$, 40, and 80~V.
 Note the drop of $E(y)$ due to the high $n^+$-doping at $y=0$, and the value of $\approx 2$~kV/cm at $y = 200~\upmu$m due to the diffusion of electrons from the $p^+$-doping into the $p$-doped Si-bulk.
 These effects are not taken into account in the analytical calculation.
 In addition, there are diffusion effects at the transition from the depleted to the non-depleted region, which however can only be seen in a figure with logarithmic $E$-scale.

 To calculate $E_{w}$, the method discussed in Sect.~\ref{sect:Ew} and Eq.~\ref{equ:Ew} with $V_0 = 1$~V is used.
 The voltage at $y=0$ is ramped linearly in time in 50~ps from $V$ to $V +1$~V, and the electrode at $y = d$ is kept on ground potential.
 The $y$-dependencies of $E_{w}(V,y,t)$ in 1~ns time steps at $V = 20$ and 80~V are shown as lines in Fig.~\ref{fig:Etw-y}, and the $t$-dependencies in the depleted region ($y = 20~ \upmu$m) and in the non-depleted region ($y = 180~ \upmu$m) as symbols in Fig.~\ref{fig:Etw-t}.
 Overall, the agreement between the TCAD simulation and the analytical calculation is very good.
 As expected, differences occur at $y=0$, at $y=w$ and close to $y = 200~\upmu$m.

 It is concluded that the analytical formula, Eq.~\ref{equ:Ewmodel}, can be used to calculate the time-dependent weighting field in non-irradiated pad diodes, and that the proposed method using TCAD simulations to calculate  time-dependent weighting fields gives reliable results.

  \section{Weighting field of the irradiated pad diode}
  \label{sect:Ewirr}

 For the simulation of the irradiated pad diode the \emph{Hamburg Penta Trap Model} HPTM~\cite{Schwandt:2018} is used.
 The HPTM assumes 5 traps, 2 acceptors and 3 donors, with energies in the band gap taken from microscopic measurements.
 The introduction rates and the cross-sections for holes and electrons have been obtained by minimising the sum of the squares of the relative differences of simulations and measurements from pad diodes irradiated with 24~GeV/c protons to $\Phi _{eq} = $~0.3, 1.0, 3.0, 6.0, 8.0 and $13 \times 10^{15}$~cm$^{-2}$ and annealed at 60~$^\circ $C for 80~minutes.
 The experimental data used for the optimisation are \emph{I--V-, C--V-} and charge-collection-measurements with near-infrared light at $- 30$ and $-20~^\circ $C.
 The data are  well described by the simulations.

 \begin{figure}[!ht]
   \centering
   \begin{subfigure}[a]{0.5\textwidth}
    \includegraphics[width=\textwidth]{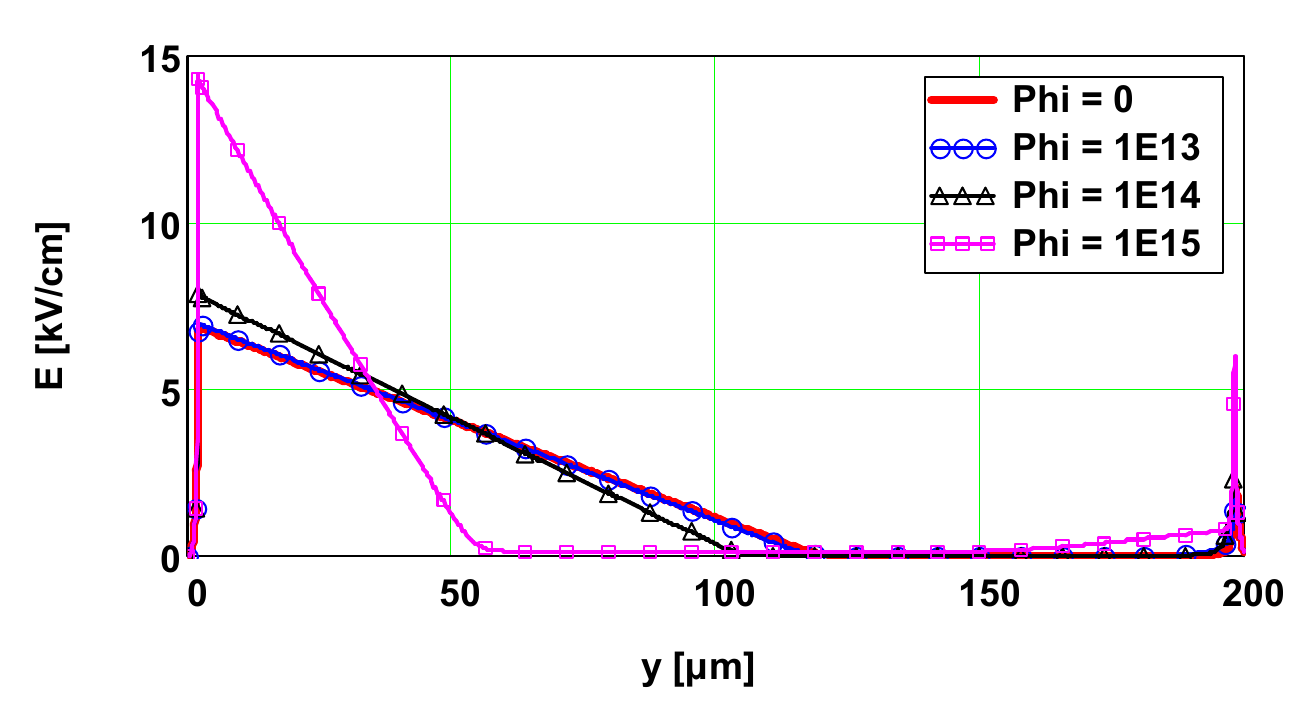}
    \caption{ }
    \label{fig:LinEfield-irr}
   \end{subfigure}%
    ~
   \begin{subfigure}[a]{0.5\textwidth}
    \includegraphics[width=\textwidth]{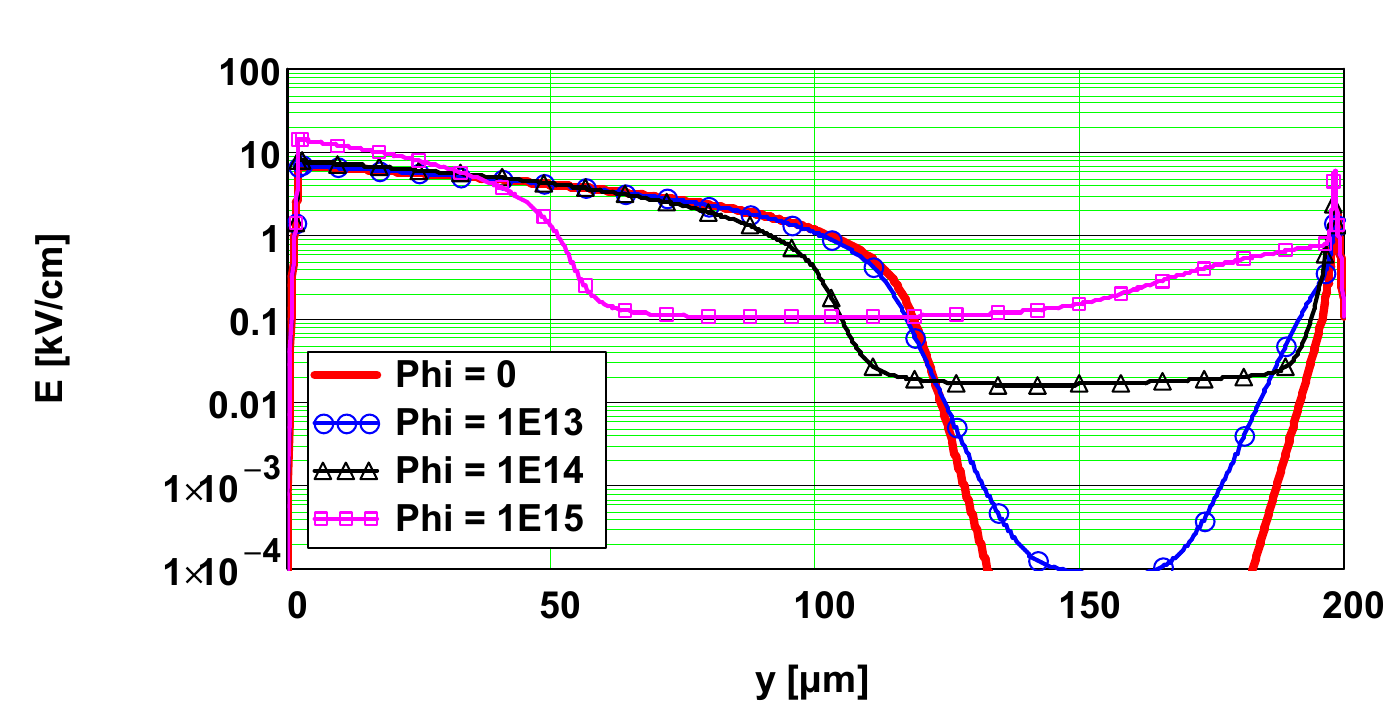}
    \caption{ }
    \label{fig:LogEfield-irr}
   \end{subfigure}%
   \caption{TCAD simulation of the electric field at $V = 40 $~V for different $\Phi _{eq}$ in units of cm$^{-2}$.
     (a) Linear, and
     (b) logarithmic scale.
    }
  \label{fig:Efield-irr}
 \end{figure}

 The TCAD simulations are performed for $\Phi _{eq} = $~0, 0.1, 0.25, 0.5, 0.75, 1, 10, and $100 \times  10^{13} $~cm$^{-2}$.
 Fig.~\ref{fig:Efield-irr} shows the electric field $E(y)$ at $V = 40$~V for selected $\Phi _{eq}$-values.
 Compared to the situation before irradiation, $E(y)$ hardly changes in the depletion region up to $\Phi _{eq} = 10^{13}$~cm$^{-2}$, whereas in the non-depleted region a small field appears.
 It is the result of the ohmic voltage drop given by the product of dark current and silicon resistance.
 The resistance  increases with $\Phi _{eq}$ because of the increase in generation rate, which results in a decrease of the density of majority charge carriers, in order to satisfy the equilibrium condition $n_e \cdot n_h =n_i^2$.
 The electron, hole and intrinsic densities are denoted $n_e$, $n_h$ and $n_i$, respectively.
 For higher $\Phi _{eq}$-values, $E(y)$ increases significantly at low $y$-values because of the filling of radiation-induced donor traps with electrons from the dark current.
 At $\Phi _{eq} = 10^{15}$~cm$^{-2}$ there is also an evidence for the increase of $E(y)$ close to $y = d$, known as \emph{double junction} in the literature~\cite{Eremin:2002}, due to the filling of radiation-induced acceptor traps with holes.
 At the voltage of 40~V the $E(y)$-field is low in a significant fraction of the pad diode.
% It is caused by the dark current and the ohmic resistance of the radiation-damaged silicon bulk.
  \begin{figure}[!ht]
   \centering
   \begin{subfigure}[a]{0.5\textwidth}
   \centering
    \includegraphics[width=\textwidth]{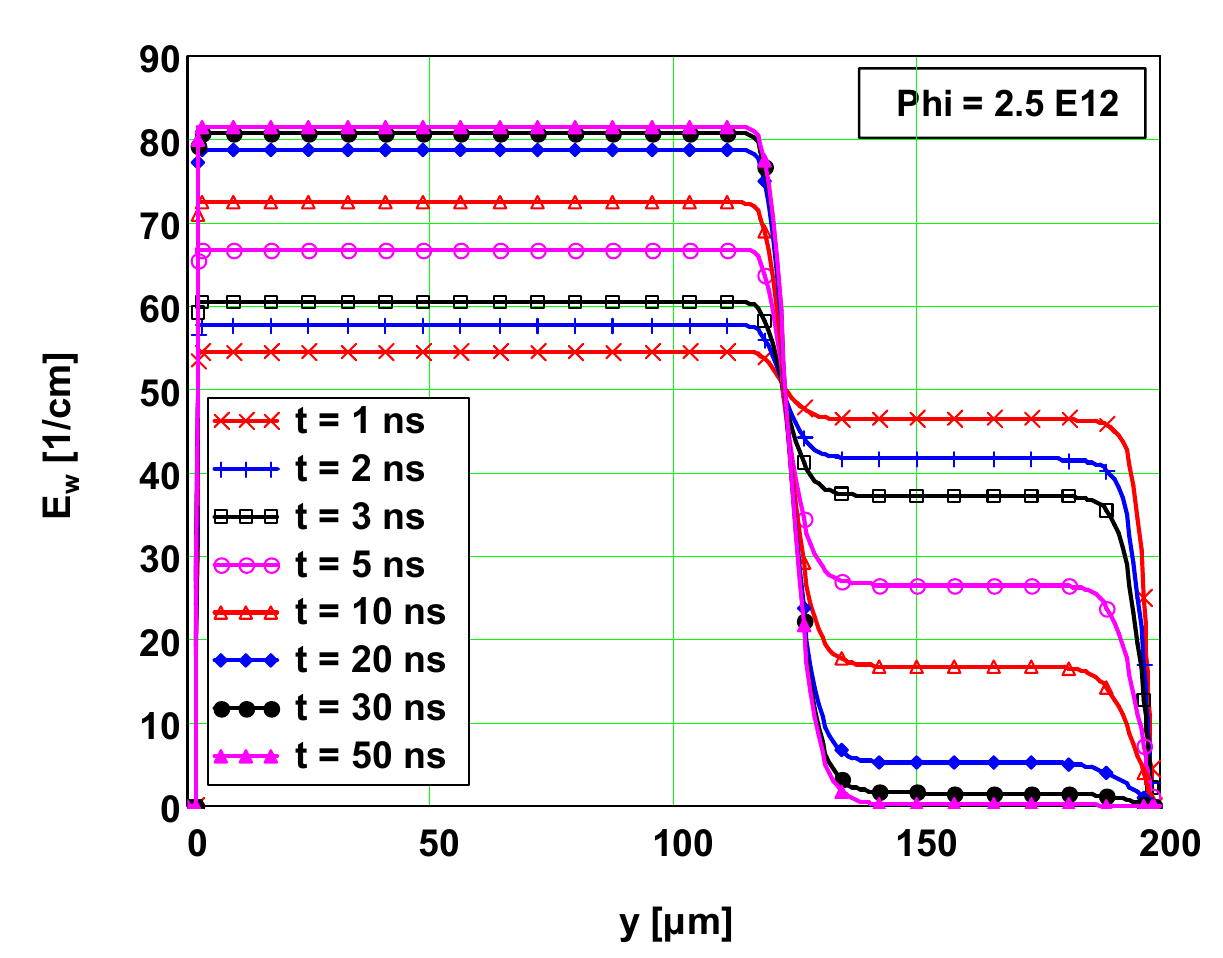}
%    \caption{ }
    \label{fig:Etw-t-2p5E12}
   \end{subfigure}%
    ~
   \begin{subfigure}[a]{0.5\textwidth}
    \includegraphics[width=\textwidth]{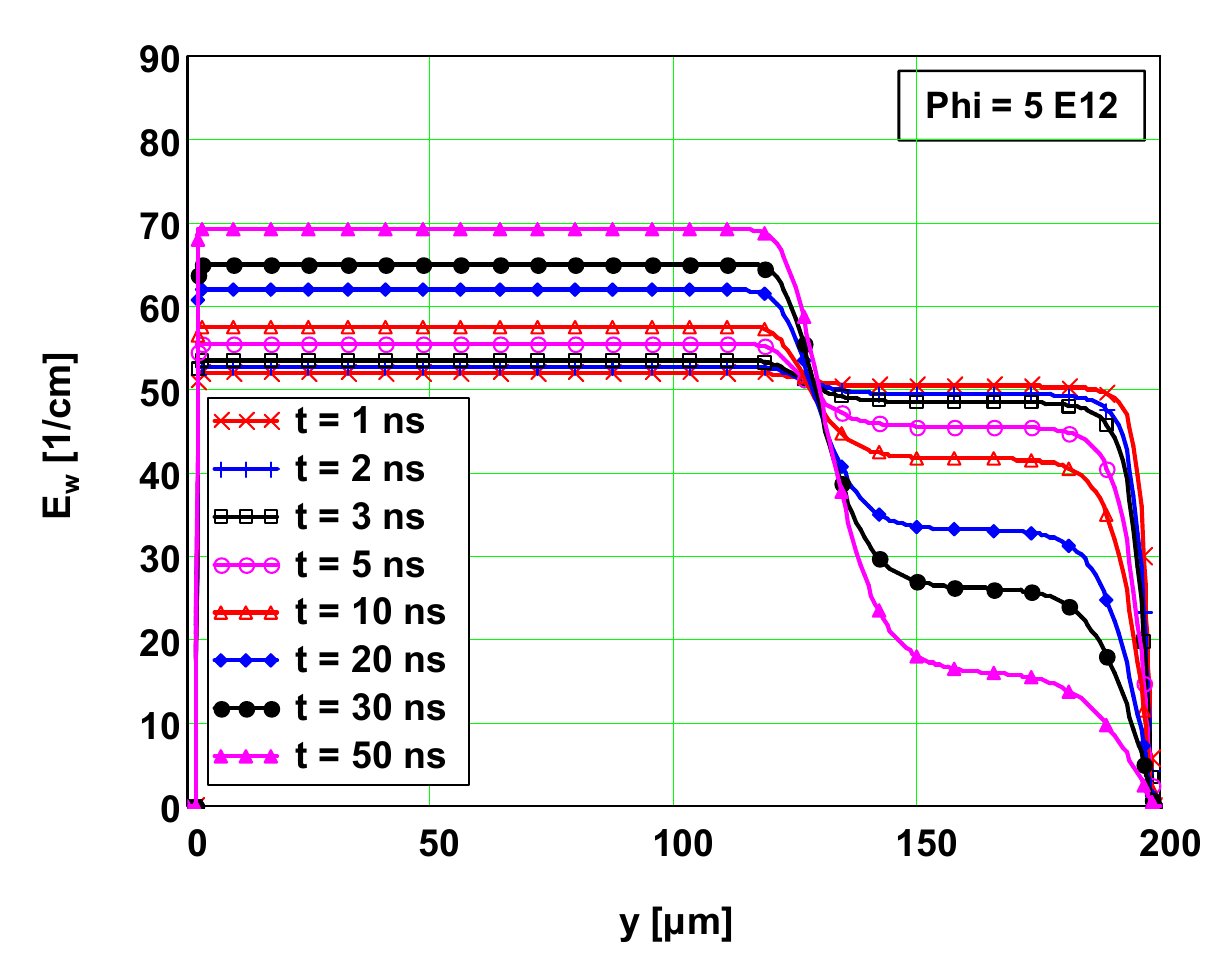}
    \label{fig:Etw-t-5E12}
%    \caption{ }
   \end{subfigure}
   ~
   \begin{subfigure}[a]{0.5\textwidth}
   \centering
    \includegraphics[width=\textwidth]{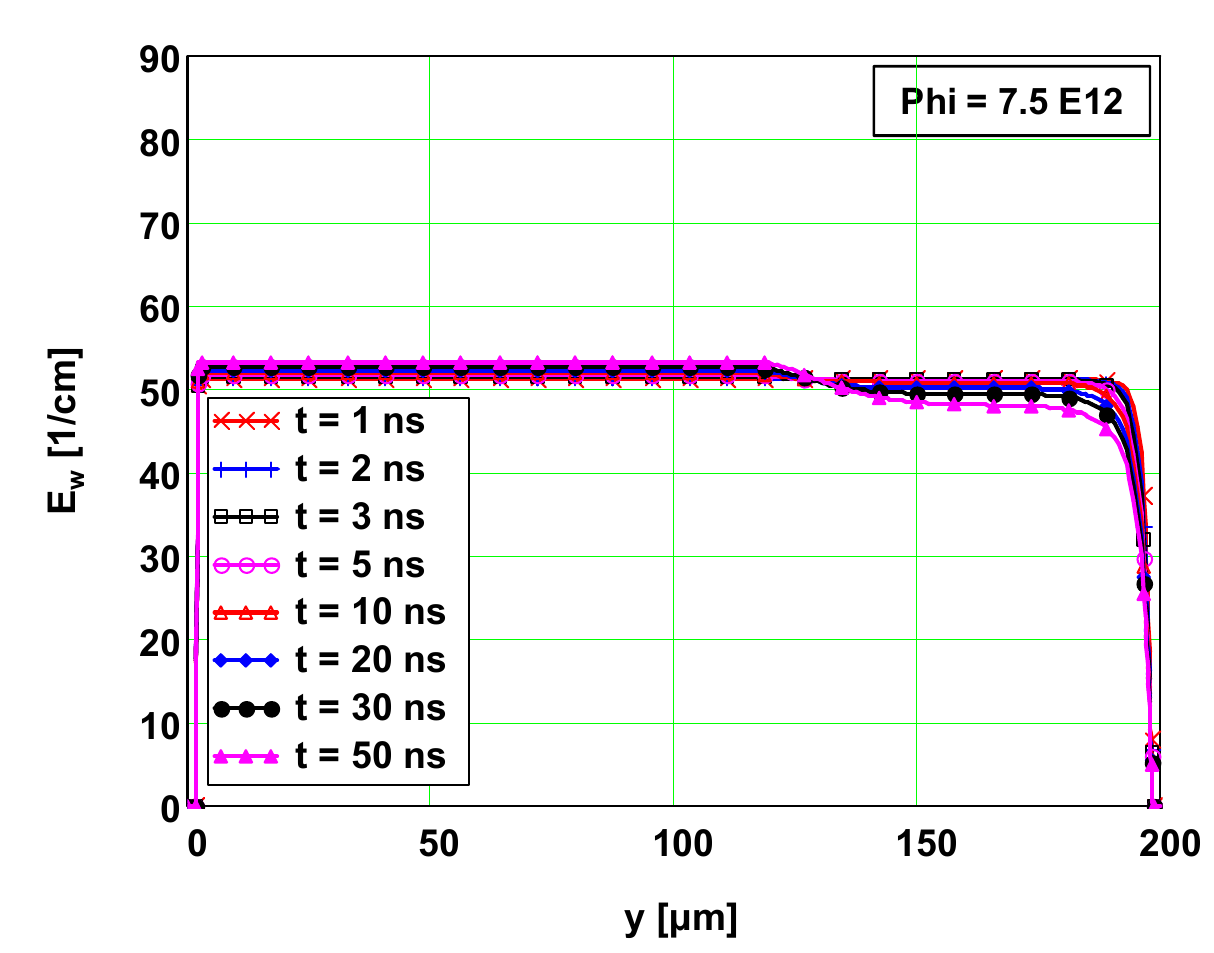}
    \label{fig:Etw-t-7p5E12}
%    \caption{ }
   \end{subfigure}%
    ~
   \begin{subfigure}[a]{0.5\textwidth}
    \includegraphics[width=\textwidth]{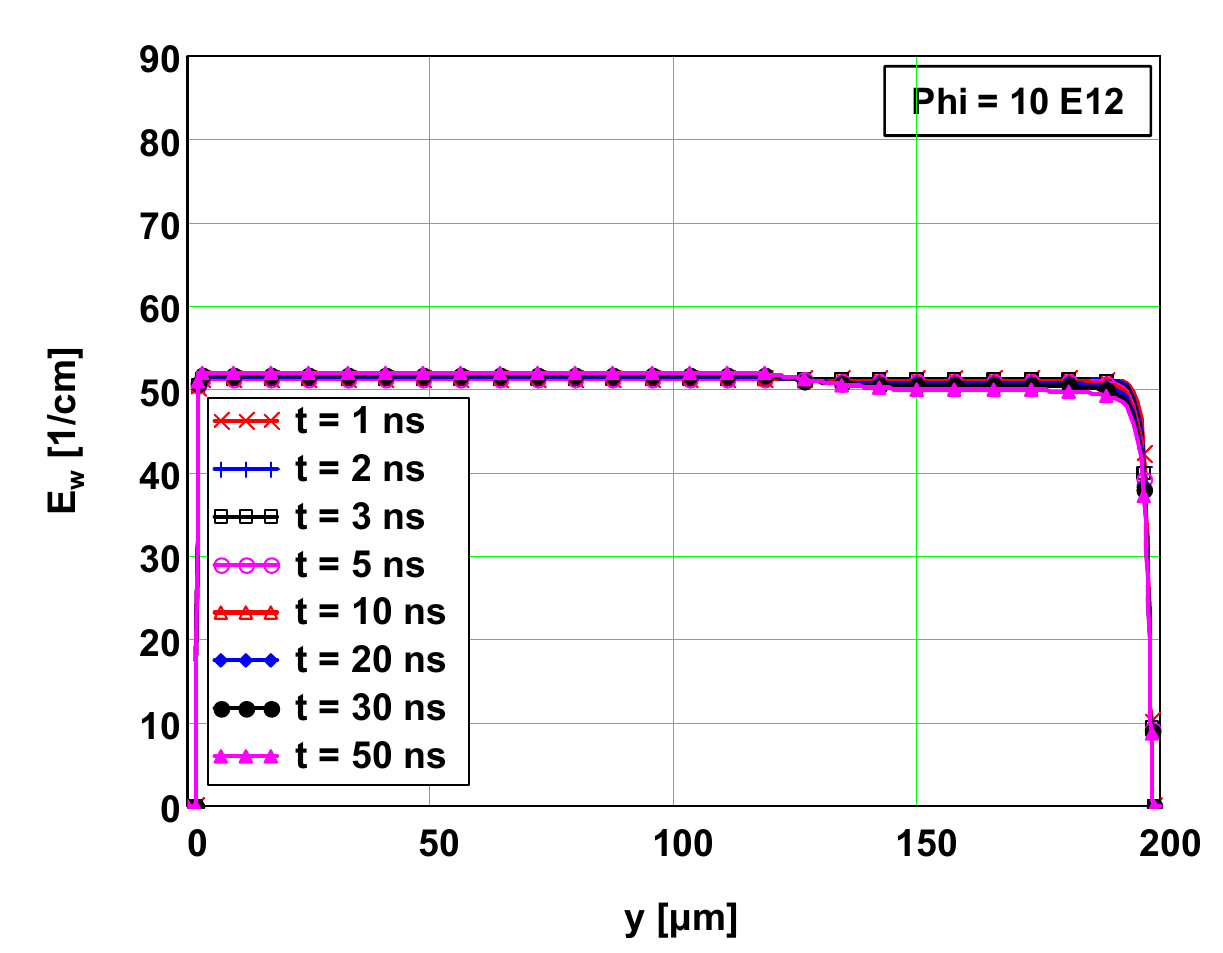}
    \label{fig:Etw-t-1E13}
%    \caption{ }
   \end{subfigure}%
   \caption{ Simulated weighting fields, $E_{w,t}$, as a function of $y$ for $t =$ ~1, 2, 3, ... 50~ns and different $\Phi _{eq} $-values given in units of cm$^{-2}$ at $V = 40$~V and $-20~^\circ$C.
   The curves closest to $1/d = 50$~cm$^{-1}$ correspond to $t = 1$~ns .
   }
  \label{fig:Etwt0}
 \end{figure}

\begin{figure}[!ht]
   \centering
    \includegraphics[width=\textwidth]{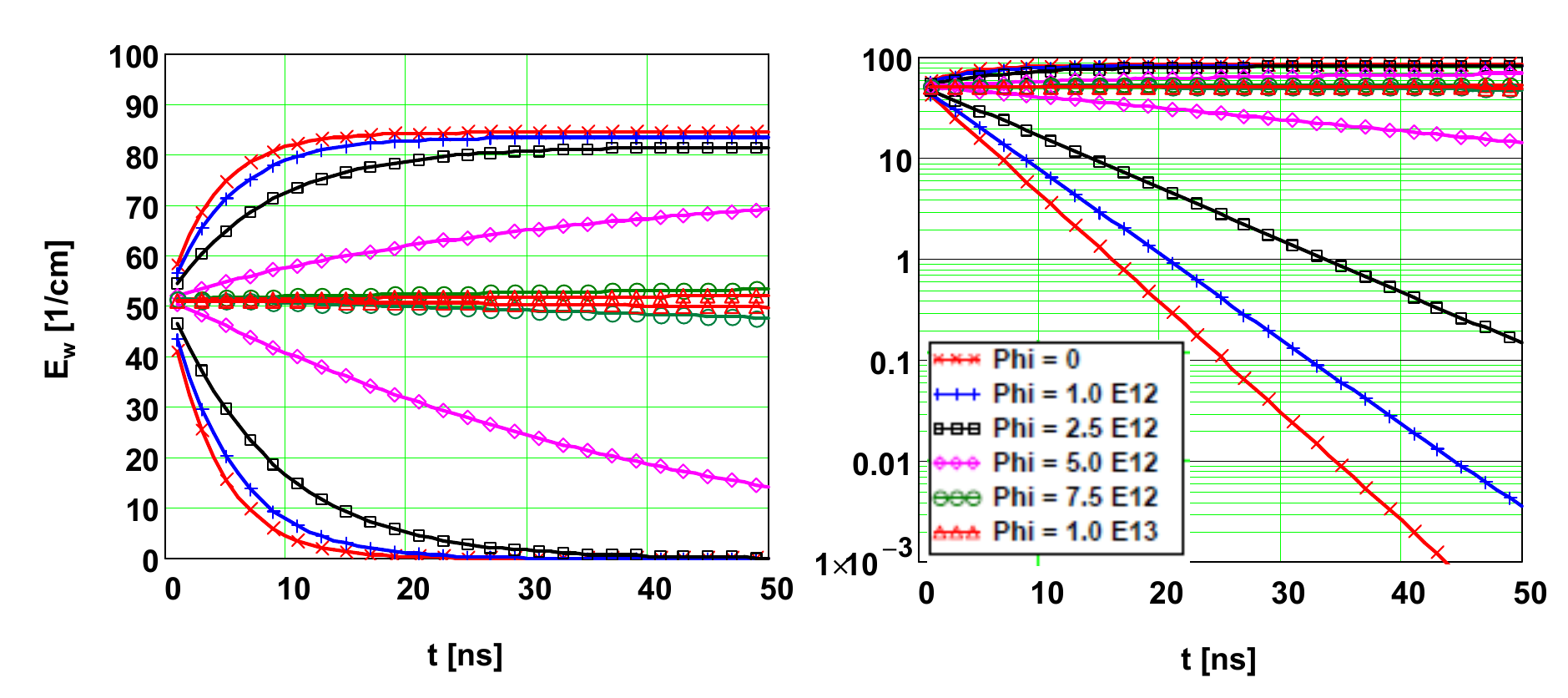}
   \caption{Time dependence of the simulated weighting field as a function of $\Phi _{eq}$ at $V = 40$~V at $-20~^\circ$C.
     The upper curves are for $y = 10~\upmu$m, the lower ones for $y = 180~\upmu$m.
     Left: linear, and right logarithmic scale. }
  \label{fig:Etw-t-Phi}
 \end{figure}

 The $y$- and $t$-dependencies of $E_w (y,t)$ from the TCAD simulation of the  pad diode at 40~V for selected $\Phi _{eq}$-values are shown in Figs.~\ref{fig:Etwt0} and \ref{fig:Etw-t-Phi}.
 Up to $\Phi _{eq} = 10^{12}$~cm$^{-2}$, the same $y$- and $t$-dependence is found as for the non-irradiated pad diode.
 Between $\Phi _{eq} = 10^{12}$ and $10^{13}$~cm$^{-2}$ the situation changes:
 The time constant of the change of $E_w(y,t)$ increases, and above $10^{13}$~cm$^{-2}$, $E _{w}(y,t)$ reaches the constant va1ue of $1/d$, which is the \emph{geometric weighting field}.
 It has been verified by TCAD simulations that $E _{w}(y,t) = 1/d$ for $t > 0$ is also valid at higher voltages if $\Phi _{eq} \gtrsim 10^{13}$~cm$^{-2}$.
 This is the justification for using the geometrical weighting field for the calculation of the current signals in highly-irradiated segmented sensors.
 It is noted that the fluence at which $E _{w}(y,t)$ becomes a constant is about one order of magnitude lower than the fluence at which the $y$-dependence of the electric field changes, indicating that these effects have different causes.

\begin{figure}[!ht]
   \centering
    \includegraphics[width=0.6\textwidth]{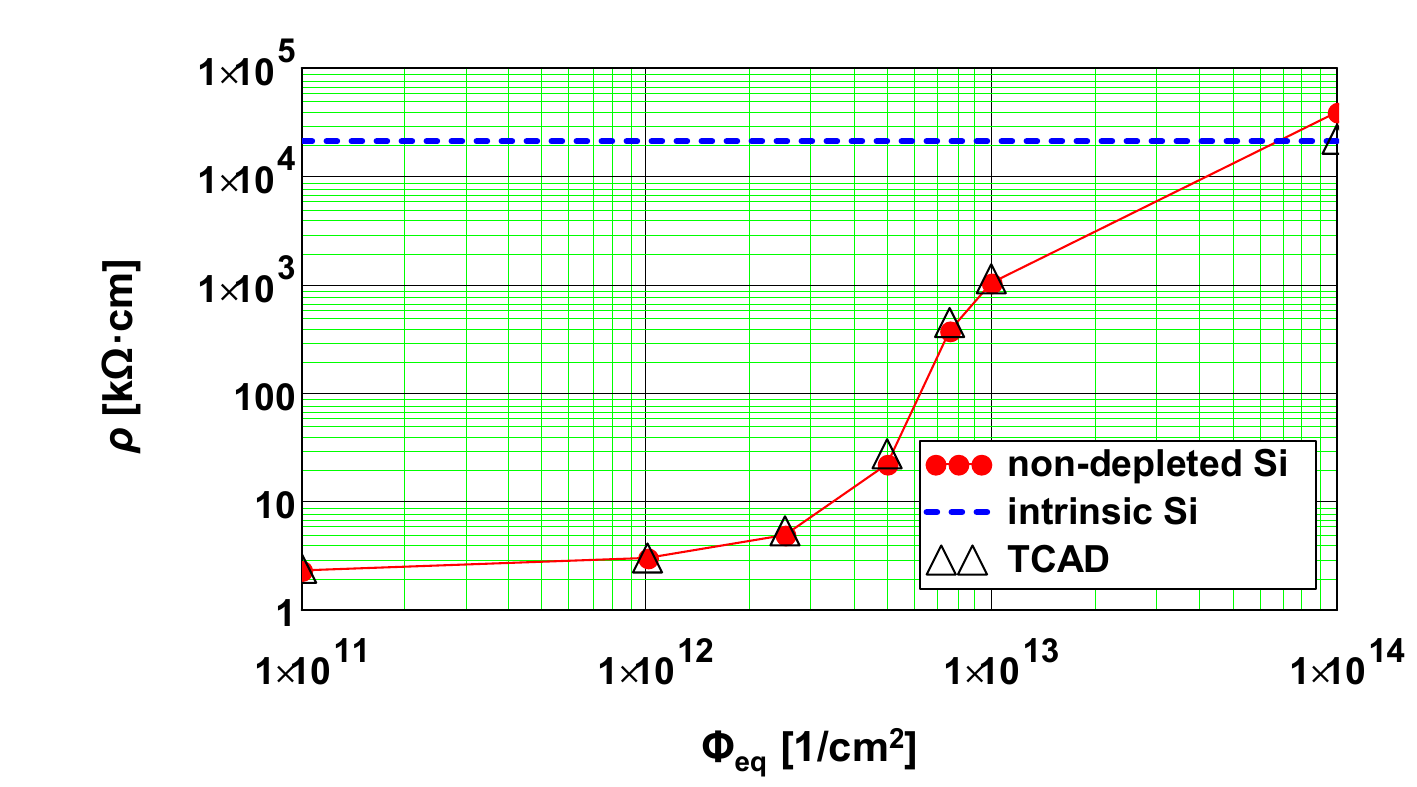}
   \caption{Full circles: Resistivity, $\rho $, in the non-depleted region for $V = 40$~V calculated from the logarithmic slope of the time-dependence of $E _{w}$ shown in Fig.~\ref{fig:Etw-t-Phi} as a function of $\Phi _{eq}$.
   Open triangles: Values of $\rho $ obtained from the TCAD simulation.
   For the calculation of $\rho $ a constant depletion width $w(40~\mathrm{V})$ is assumed.
   The value before irradiation is displayed at $\Phi _{eq} = 10^{11}$~cm$^{-2}$.
   The dashed line corresponds to the intrinsic resistivity at $- 20~^\circ$C, the temperature at which the TCAD simulations are made.
   The uncertainty of $\rho $ determined from the time dependence of $E_w$ at $\Phi _{eq} = 10^{14}$~cm$^{-2}$ is about 100~\%.  The $E_w$-decay time is about $50~\upmu$s and the TCAD simulation is only made up to 50~ns. }
  \label{fig:rho}
 \end{figure}

 Fig.~\ref{fig:Etw-t-Phi} shows the $t$-dependence of $E _{w}$ at $y=10~\upmu$m (depleted region) and at $y=180~\upmu$m (non-depleted region) for different values of $\Phi _{eq}$.
 For all $\Phi _{eq}$-values exponential dependencies are observed, with time constants $\tau (V,\Phi _{eq})$, which depend on both $V$ and $\Phi _{eq}$.
 Using Eq.~\ref{equ:tau} and $w(40$~V) of the non-irradiated diode, an estimate of the resistivity $\rho (\Phi _{eq})$ of the low field region $y > w$ can be obtained from $\tau (\Phi _{eq})$.
 As shown in Fig.~\ref{fig:rho}, $\rho (\Phi _{eq})$  increases by about four orders of magnitude from  $2.3~\mathrm{k}\Omega \cdot$cm to a value of a few $ 10^{4}~\mathrm{k}\Omega \cdot$cm.

 The value expected for the non-irradiated silicon is given by the boron-doping density and the mobility of holes: $\rho = (q_0 \cdot \mu _h \cdot N_d)^{-1} = 2.3~\mathrm{k}\Omega \cdot$cm.
 If the density of free charge carriers is dominated by generation-recombination
 \begin{equation}\label{eq:rhoi}
  \rho = \big( 2 \cdot q_0 \cdot n_i \cdot \sqrt{\mu _e \cdot \mu_h} \big)^{-1},
 \end{equation}
  with the value of $\rho = 2.1 \times 10^4~\mathrm{k}\Omega \cdot$cm for silicon at $-20~^\circ$C.
 Fig.~\ref{fig:rho} shows that the analysis of the simulated data is in agreement with these expectations.
 The mobilities of holes and electrons are $\mu _h$ and $\mu _e$, respectively, and $n_i$, $n_h$ and $n_e$ are the intrinsic charge-carrier density, and the densities of holes and electrons.
 Eq.~\ref{eq:rhoi} follows from the relation $\rho = \big( q_0 \cdot (n_h \cdot \mu _h + n_e \cdot \mu _e) \big)^{-1}$.
 Equal electron-hole generation gives $\mu _h \cdot n_h = \mu _e \cdot n_e$, the equilibrium condition is $n_h \cdot n_e = n_i^2$, from which follows $\mu _h \cdot n_h + \mu _e \cdot n_e = 2 \cdot n_i \cdot \sqrt{\mu _e \cdot \mu _h}$.

% For the highly-irradiated silicon  the high-generation/annihilation rate results in $n_e = n_h = n_i$, and its intrinsic resistivity $\rho _{i} = \big( q_0 \cdot n_i \cdot (\mu_h + \mu_e) \big)^{-1} \approx 1.5 \times 10^4~\mathrm{k}\Omega \cdot $cm.
% The mobilities of holes and electrons are $\mu _h$ and $\mu _e$, respectively, and $n_i$, $n_h$ and $n_e$ are the intrinsic charge-carrier density, and the densities of holes and electrons.
% That the value of $\rho $ is even large than $\rho _{i}$ for $\Phi _{eq} = 10^{15}$~cm$^{-2}$, is ascribed to the assumption that the depth of low-field region is $w(40$~V), which not correct, as can be seen from Fig.~\ref{fig:Efield-irr}.

 To summarise this section:
 The TCAD simulations show that for $\Phi _{eq}$ above $10^{13}$~cm$^{-2}$ the weighting field is independent of time and voltage, and equal to the \emph{geometric weighting field}, i.~e. the time-inde\-pendent weighting field of the fully depleted non-irradiated sensor.
 At lower voltages and fluences, the time dependence of $E_{w}$ is related to the resistivity of the low-field region. Its resistivity increases with $\Phi _{eq}$ and  reaches approximately the intrinsic resistivity.
% Given the relation between the time dependence of the weighting field and the frequency dependence of the sensor capacitance,  it can be concluded that the frequency dependence of the capacitance for irradiated sensors at low voltages is mainly caused by the resistivity of the silicon in the low-field region and not by the detailed properties of the radiation-induced traps in the silicon band gap.
% Above a certain frequency, the capacitance of an irradiated sensor becomes independent of voltage and frequency, and its value is the value of the capacitance of the fully depleted, non-irradiated sensor.

\newpage

  \section{Conclusions}
  \label{sect:Conclusions}

 In this paper the weighting field, $E_w$, is investigated, which is needed to simulate the response of radiation-damaged silicon detectors.
 Usually it is assumed that $E_w$ for irradiated sensors does not depend on time and can be calculated as the difference of the electric field in the biased sensor and 1~V added to the readout electrode  minus the electric field in the biased sensor.
 The paper shows that this assumption is valid, and Eq.~\ref{equ:Igeom} can be used to calculate the induced current.

 As pointed out in Ref.~\cite{Riegler:2018}, a time-dependent $E_w$ is required to simulate the response of partially depleted, non-irradiated silicon sensors.
 Using TCAD simulations of a pad diode with radiation damage described by the Hamburg Penta Trap Model~\cite{Schwandt:2018}, it is shown that for a partially depleted sensor and neutron equivalent fluences $\Phi _{eq} \lesssim 10^{13}$~cm$^{-2}$ a time-dependent $E_w$ is required, whereas at higher fluences $E_w$ is time independent.
 The reason for the transition is the increase of the resistivity in the low-field region of the sensor due to the increase of the charge-carrier generation rate by radiation damage.
 At $ T = - 20~^\circ $C, the temperature of the study, the resistivity of the non-depleted silicon increases by about four orders of magnitude as a function of $\Phi _{eq}$.

% \section*{Acknowledgements}
% \label{sect:Acknowledgement}

%\newpage

\end{document}